\documentclass[11pt]{article}
\usepackage{amsmath,amsthm,latexsym,amssymb,amsfonts,parskip,epsfig}
\usepackage[a4paper,top=2cm,bottom=3cm]{geometry}
  
\numberwithin{equation}{section}
\allowdisplaybreaks

\newtheorem{lemm}{Lemma}[section]
\newtheorem{prop}[lemm]{Proposition}

\newcommand{\R}{\mathbb{R}}                  
\newcommand{\C}{\mathbb{C}}                  

\newcommand{\oiint}{\bigcirc \hspace{-1.35em} \int \hspace {-0.8em} \int}

\newcommand{\norm}[1]{\left\lVert #1 \right\rVert}
\newcommand{\betr}[1]{\left\lvert #1 \right\rvert}

\newcommand{\expec}[1]{\langle #1 \rangle}
\newcommand{\dm}{\text{d}\mu}

\newcommand{\mal}{\mu_{\text{AL}}}
\newcommand{\pexp}[1]{\mathcal{P}\exp \oint_{#1}}
\newcommand{\ppexp}[1]{\mathcal{P}\exp \oiint_{#1}}

\newcommand{\wh}[1]{\widehat{#1}}
\newcommand{\lieg}{\mathfrak{g}}

\DeclareMathOperator{\abar}{\overline{\mathcal{A}}}
\DeclareMathOperator{\conn}{\mathcal{A}}
\DeclareMathOperator{\tr}{tr}

\DeclareMathOperator{\ad}{ad}
\DeclareMathOperator{\SUTWO}{SU(2)}
\DeclareMathOperator{\SUTHREE}{SU(3)}
\DeclareMathOperator{\SL2C}{SL(2,\C)}

\DeclareMathOperator{\one}{1}
\DeclareMathOperator{\um}{\mathbb{I}}

\DeclareMathOperator{\sym}{Sym}

\title{Chern-Simons theory, Stokes' Theorem, and the Duflo map}
\author{Hanno Sahlmann$^*$\\ Asia Pacific center for Theoretical Physics, Pohang (Korea)\\[.2cm]
Thomas Thiemann$^\dagger$\\ Institute for Theoretical Physics III, Erlangen University (Germany)}
\date{{\small Preprint APCTP Pre2010-007}}

\begin{document}
\maketitle
\begin{abstract}
We consider a novel derivation of the expectation values of holonomies in
Chern-Simons theory, based on Stokes' Theorem and the functional properties of the Chern-Simons action. It involves replacing the connection by certain functional derivatives under the path integral integral. 
It turns out that ordering choices have to be made in the process, and we demonstrate that, quite surprisingly, the Duflo isomorphism gives the right ordering, at least in the simple cases that we consider. In this way, we determine the expectation values of unknotted, but possibly linked, holonomy loops for SU(2) and SU(3), and sketch how the method may be applied to  more complicated cases.  
Our manipulations of the path integral are formal but well motivated by a rigorous calculus of integration on spaces of generalized connections which has been developed in the context of loop quantum gravity.  
\end{abstract}
\section{Introduction}
{\renewcommand{\thefootnote}{\fnsymbol{footnote}}
\footnotetext[1]{sahlmann@apctp.org}
\footnotetext[2]{thiemann@theorie3.physik.uni-erlangen.de}
}
Chern-Simons (CS) theory is the prototypical example of a
topological (quantum) field theory. It is of great interest
both in mathematics and in physics. On the mathematical side
has applications to knot theory, topology of three dimensional
manifolds, and quantum groups among many others.
For physics it is of interest because -- as a topological
theory -- it is a model for the quantization of background
independent theories, most importantly for gravity. Even more,
Chern-Simons theory \textit{is} Einstein gravity in three
dimensions for specific choices of the structure group
(depending on the signature and the existence and sign of
the cosmological constant).\footnote{
Additionally there may be an important link between
CS theory and the
quantization of gravity in four dimensions: It
has been argued that the exponential of the action for CS
theory with $G=\SL2C$ is a solution to all the constraints
that appear in the canonical formulation of 4d gravity
(with cosmological constant). This state is known as the
Kodama state \cite{Kodama:1990sc,Freidel:2003pu}. Whether it
represents a physically viable state is however in dispute
\cite{Freidel:2003pu,Witten:2003mb}.}
In the present article, we want to describe a new procedure to calculate expectation values in non-abelian CS theory. In a companion article \cite{Sahlmann:2010bd}, we have described how reasoning along similar lines can be used to completely determine the expectation values in the case of abelian CS theory. Since CS theory has been studied very extensively, and expectation values can be computed easily and explicitly for a large variety of structure groups, our goal is not that of again determining these well known quantities. 
Rather, we hope that the new approach can perhaps illuminate some new aspects of the theory. In particular, we see that the Duflo-isomorphism seems to play an important role. A similar observation has been made before, but in a seemingly unrelated context \cite{Freidel:2005ec}.   

Our approach is very similar in spirit to the pioneering work \cite{Gambini:1996mb,Gambini:1997fn}, in that it uses Stokes' theorem, and functional derivatives. It does differ in other aspects, most notably that we will not work with infinitesimal loops, and will not focus on the skein relations. Also the use of the Duflo isomorphism is new. 

\begin{figure}
\centerline{\epsfig{file=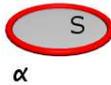, scale=0.35}}
\caption{\label{fi_unknot_seifert} A simple Wilson loop
$\alpha$ and a surface $S$ bounded by it}
\end{figure}

With that said, let us describe the idea in more detail. Stokes' Theorem relates the contour integral
of an Abelian  connection $A$ around a loop $\alpha$
bounding a surface $S$ (as in figure
\ref{fi_unknot_seifert}) to the integral of the curvature
over the surface $S$. 
For the non-abelian case there exists a generalization,
\begin{equation}
\pexp{\alpha}A\, \text{d}s=\ppexp{S} \mathcal{F}\,
\text{d}^2 s.
\end{equation}
Here $\mathcal{F}=hFh^{-1}$ is a
certain parallel transport of the curvature $F=D_A A$ of
$A$, and the surface integral on the right hand side is
\textit{surface ordered}. It is important to note that for this generalization, $S$ must be simply connected. Using Stokes' Theorem, one can thus replace a holonomy functional depending on the connection, with a functional depending on the corresponding curvature.  
A basic property of the CS action is 
\begin{equation}
\frac{\delta}{\delta A(x)} S_{\text{CS}}[A]\propto F(x), 
\end{equation}
thus when working under the path integral of CS theory, one can formally replace curvature with functional differentiation (see also \cite{Gambini:1996mb}). Taking the two transformations together one can replace a holonomy functional under the path integral with a complicated functional differential operator. This differential operators acts on the exponentiated action, and it may also act on other holonomy functionals that are present under the integral. It may even act on itself, in the case that the path of the holonomy has transversal intersections with the surface $S$ that it bounds. The action on the exponentiated CS action can be transferred
to the other holonomy loops by a partial integration. Then, if the action of the differential operator can be evaluated, one obtains a new representation of the expectation value that one started with. As an example, for the product of two holonomy functionals $W_\alpha,W_\beta$, one can obtain 
\begin{equation}\begin{split}
\expec{W_\beta W_\alpha}&=\int_{\conn} W_\beta[A]W_\alpha[A] \exp(i
S_{\text{CS}}[A])\, \dm [A]\\
&=\int_{\conn}W_\beta\tr \left[\ppexp{S} \mathcal{F}\,
\text{d}a\right]\exp(i S_{\text{CS}}[A])\, \dm [A]\\
&=\int_{\conn}W_\beta\tr \left[\mathcal{P}\exp
\frac{2\pi c}{k}\oiint_{S} h_x^{-1}
\overrightarrow{\frac{\delta}{\delta A}}
h_x\text{d}a\right]
\exp(i S_{\text{CS}}[A])\, \dm [A]\\
&=\int_{\conn}W_\beta\tr \left[\mathcal{P}\exp
-\frac{2\pi c}{k}\oiint_{S} h_x^{-1}
\overleftarrow{\frac{\delta}{\delta A}} h_x\text{d}a\right]
\exp(i S_{\text{CS}}[A])\, \dm [A]\\
&=\expec{\boldsymbol{W_{S}}^\dagger[W_\beta]}.
\end{split}\end{equation}
Here, $\conn$ is a suitable space of connections, $\dm [A]$ a Lebesgue-like measure, and the derivations are assumed to be anti-symmetric with respect to the latter. $\boldsymbol{W_{S}}[\ldots]$ denotes the action of the functional differential operator obtained from $W_\alpha$.

We are motivated to consider this type of transformation by the fact that many of the ingredients needed are actually available in a rigorous setting developed for loop quantum gravity. In particular:  
\begin{itemize}
\item[(a)] We denote by $\abar$ the space of generalized $G$-connections, with $G$ a compact Lie group (see for example \cite{Ashtekar:1994mh}).
This is a space of distributional connections. It is compact, Hausdorff, and 
smooth connections are dense.

\item[(b)] We denote by 
$\mal$ the Ashtekar-Lewandowski measure \cite{Ashtekar:1994mh}, a non-degenerate, uniform measure on the space $\abar$. 

\item[(c)] A rigorous definition of the functional
derivative $\delta/\delta A$ has been given \cite{Ashtekar:1996eg}. More precisely:
\begin{equation}
\label{eq:der}
\wh{X}_{S}=\int_S \epsilon_{cab}\frac{\delta}{\delta A_c}\text{d}x^a\wedge
\text{d} x^b
\end{equation}
for any surface $S$ gives a well defined derivation on suitably differentiable
holonomy functionals on $\abar$. We will describe its action in more detail below. Here we only mention that it has the expected
adjointness properties with respect to the scalar product induced by $\mal$:
\begin{equation}
\wh{X}_{S}^\dagger = - \wh{X}_{S}.
\end{equation}
\end{itemize}
What is more, in the examples that we will consider, the action of the operators $\boldsymbol{W_{S}}$ are well defined and map holonomy functionals onto holonomy functionals. This results in a simplified expression for the original expectation value. 

Some remarks are in order: 

First of all, in reality, the CS path integral measure is certainly not the product of an integrable function with a Lebesgue type measure, as suggested above. Rather, the heuristic procedure sketched above will, when implemented in detail, provide results for the expectation values which, taken together, define the path integral measure. 

\begin{figure}
\centerline{\epsfig{file=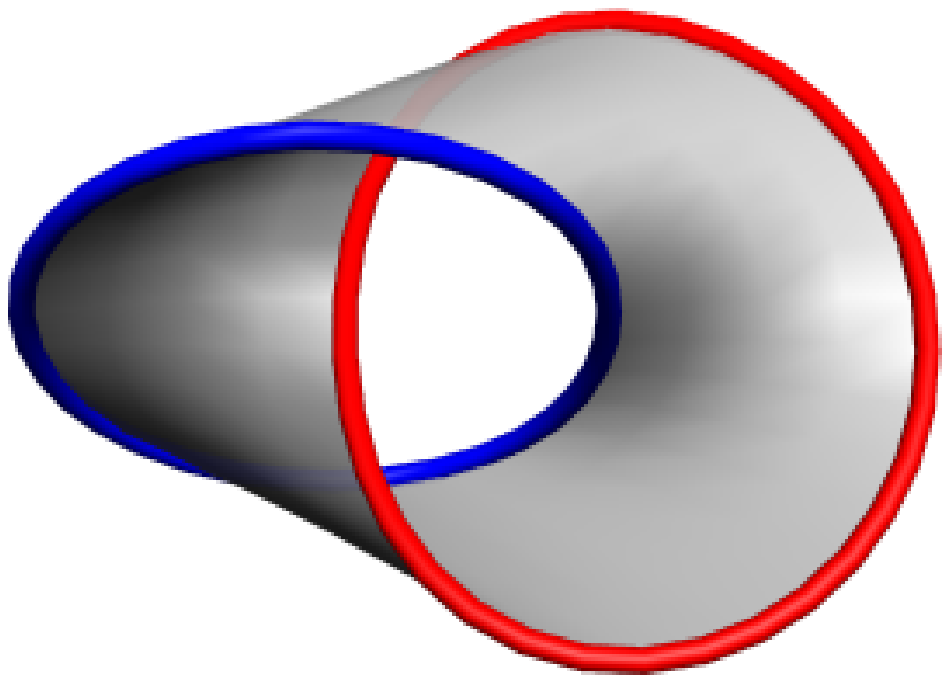, scale=0.33}
\epsfig{file=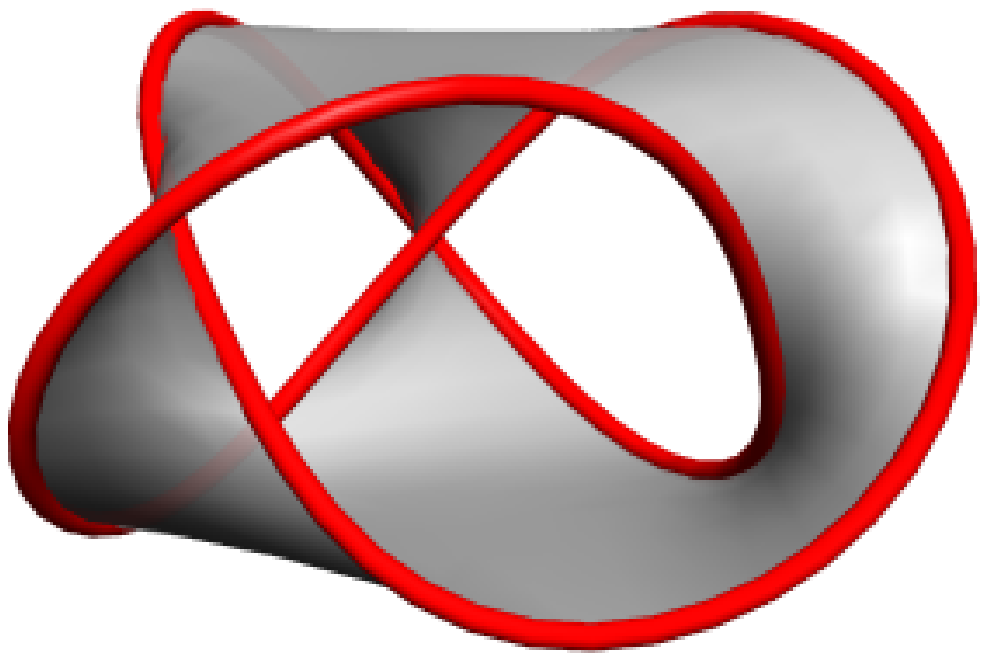, scale=0.33}}
\caption{\label{fi_seifert_example} Examples for Seifert
surfaces of knots and links: For the Hopf rings (left)
and the trefoil knot  (right). Graphics here and in some of the following figures created with \texttt{SeifertView} \cite{SV})}
\end{figure}

Secondly, the procedure requires the choice of surfaces with the holonomy loops as boundaries. Moreover, for the application of the non-abelian generalization of Stokes' Theorem, these surfaces have to be simply connected. The question of existence of such surfaces can not be answered independently of the manifold $M$ into which the loops are embedded. We will restrict ourselves here to $M=S^3$. In this case, there always exist connected and orientable surfaces that bound a given loop, the so-called \emph{Seifert surfaces} (see figure \ref{fi_seifert_example} for two examples). 
While these were sufficient for the abelian case \cite{Sahlmann:2010bd}, here they are not, due to the fact that they are in general not simply connected. We will discuss this problem further, below. 
As to the question of uniqueness of these surfaces, we remark that in the abelian case, it turned out that all that mattered about the surface was the framing that it induced for the loop. Thus expectation values were unique up to framing. But this framing dependence is precisely what one expects for CS theory, both in the abelian and in the non-abelian case. Thus we strongly suspect that the choice of surface corresponds to the choice of framing again in the non-abelian case. In fact, we find out that this is true in most of the cases that we consider.

Finally, it turns out that the functional derivatives \eqref{eq:der} do not commute with each other. This is well known \cite{Ashtekar:1998ak,Baratin:2010nn}, and leads to non-commutativity of spatial geometry in loop quantum gravity. There is thus an ordering ambiguity in the operators $\boldsymbol{W_{S}}$. All the difficulty in the approach lies with determining the correct ordering prescription for the functional derivatives. It is here that the Duflo isomorphism \cite{duflo} enters the picture. This isomorphism is a map between the symmetric algebra $\sym(\lieg)$ and the universal enveloping algebra $U(\lieg)$ of a given Lie algebra $\lieg$, which is an algebra isomorphism between the invariant subspaces of the algebras involved. It is, in a sense, a quantization map, and we will show that it is very well adapted to dealing with the ordering problem that we face. 

We should stress that there are many other works that apply loop quantum gravity techniques to CS theory, besides \cite{Gambini:1996mb,Gambini:1997fn} and the present one. An incomplete list is
\cite{Major:1995yz,Borissov:1995cn,Noui:2004iy,Freidel:2004vi,Freidel:2004nb,Freidel:2005bb,Meusburger:2008bs}
Much of this work is much more advanced than what we will present here. Still we think it is worthwhile to investigate the new approach because it differs in some respects from the other work. For example, we do not make any use of a discretization of the path integral, as is done in spin foam models \cite{Freidel:2004vi,Freidel:2004nb,Freidel:2005bb} and   
we do not use the framed quantum spin networks of \cite{Major:1995yz,Borissov:1995cn}.

The article is structured as follows. In the next section, we describe the approach in more detail. In particular, we discuss regularization and ordering choices. A particular aspect of these choices, the use of the Duflo isomorphism, is further explained in section \ref{se_duflo}. 
In section \ref{se_unknots} we use it to calculate the expectation values of unknots. Section \ref{se_hopf} contains the calculation of the expectation values for Hopf links and some more general links. In section \ref{se_disc}, we discuss the results and give some perspectives on generalizations. There are also two appendices. One lists some well known results on CS expectation values for comparison. The other contains some Lie algebra traces that are needed in the main text. 

\section{Surface ordering, factor ordering, regularization}
\label{se_order}
In the present section, we will make the transformations under the path integral that were sketched in the introduction more precise, and discuss the ordering and regularization issues that arise. Some notations and conventions have to be discussed first, however. 

As in the introduction, $G$ stands for a compact gauge group, and we will denote by $\lieg$ its Lie algebra. We will additionally take $G$ to be semi-simple, although it seems that much of what we are going to do would generalize.  Sometimes, $G$ will be specialized to either SU(2) or SU(3), but if not specified otherwise, calculations go through for any semi-simple $G$. We will not fix any particular basis for $\lieg$, but we use the notation $t_I$ for the elements of an (arbitrary) basis $\{t_I\}_I$ of $\lieg$. Given such a basis, we will denote by $T_I$ the matrix corresponding to $t_I$ in the defining representation of $\lieg$. Also, $\{t^I\}_I$ will denote the basis in the dual $\lieg^*$ such that $t^J(t_I)=\delta^J_I$. 

We introduce the following notation for the Cartan-Killing form on $\lieg$:
\begin{equation}
\norm{h}^2:=\tr(\ad(h)\ad(h))=:h^Ih^J k_{IJ}
\end{equation}
for $h\in \lieg$, with $h=h^I t_I$ for some basis $\{t_I\}_I$, and $\ad$ the adjoint representation. Indices will always be lowered or raised using
$k_{IJ}$ or its inverse, $k^{IJ}$. In particular we will define $T^I:=k^{IJ}T_J$, which needs to be distinguished from $t^I$, an element of the dual basis. 
In addition to the Cartan-Killing form, we also need
\begin{equation}
\betr{h}^2=\tr(\pi(h)^2)=:h^Ih^J \widetilde{k}_{IJ} 
\end{equation}
where $\pi$ is the defining representation. 
Then, for the two algebras that will interest us in the following, we have 
\begin{equation} 
 \widetilde{k}_{IJ}= c k_{IJ} \text{ with } c
 =\begin{cases}
 \frac{1}{4} &\text{ for SU(2)}\\
 \frac{1}{6} &\text{ for SU(3)}
\end{cases}.
\label{eq:c}
\end{equation}
One slight exception to the conventions laid out above occurs in section \ref{se_duflo}, where we will denote a basis element of $\lieg$ -- in its capacity as a linear functional on the dual $\lieg^*$ of $\lieg$ -- as $E_I$. In that context, we will also write $E$ to denote the matrix $E_IT^I$ of functions on $\lieg^*$.
 
With these definitions out of the way, let us now start the discussion by describing the 
non-Abelian analog of Stokes' Theorem that we will use. There exist many variants. We will use the one stated and proven in \cite{nastokes}, which we will re-state here for the convenience of the reader.  

Let a surface $S$ be parameterized by a function $S:[0,1]^2\longrightarrow \R^3$. Let the coordinates
of a point $p$ induced by this parametrization be $(s(p),t(p))$.
We define a system of
paths on $S$, all starting from the point $S(0,0)$ as follows:
\begin{equation}
w(p)=S([0,s(p)],0)\cup S(s(p),[0,t(p)])
\end{equation}
Let a connection $A$ on $S$ be given, and denote with $F$ its curvature. Then let
\begin{equation}
\mathcal{F}_{ab}(p):= h_{w(p)}F_{ab}(p)h_{w(p)}^{-1},
\end{equation}
which, as far as gauge transformations are concerned, is based at $S(0,0)$
\begin{prop}[\cite{nastokes}]
The holonomy $h_{\partial S}$ around the surface (i.e. along the path  $S(0,0)$ to $S(1,0)$ to $S(1,1)$ to $S(0,1)$ and back to $S(0,0)$) can be expressed as 
\begin{equation}
\begin{split}
h_{\partial S}=\um &+\int_0^1\text{d}s_1\int_0^1\text{d}t_1\, \mathcal{F}(S(s_1,t_1))\\
&+\int_0^1\text{d}s_1\int_0^1\text{d}t_1\int_0^{s_1}\text{d}s_2\int_0^1\text{d}t_2\, \mathcal{F}(S(s_2,t_2))\mathcal{F}(S(s_1,t_1))\\
&+\ldots\\
&=: \ppexp{S} \mathcal{F}\,
\text{d}^2 s
\end{split}
\label{eq:nastokes}
\end{equation}
\end{prop}
We want to replace the connection by functional derivatives. We use that for the Chern-Simons action 
\begin{equation}
S_\text{CS}=\frac{k}{4\pi}\int_M\tr(A\wedge dA+\frac{2}{3}A\wedge A\wedge A))
\end{equation}
it holds that 
\begin{equation}
\frac{\delta}{\delta A^I(x)} e^{iS_{\text{CS}}}[A]= \frac{ick}{2\pi} F_I(x)e^{iS_{\text{CS}}}[A]. 
\label{eq:func}
\end{equation}
Here and in the following $c$ is the constant from \eqref{eq:c}. 

\begin{figure}
\centerline{\epsfig{file=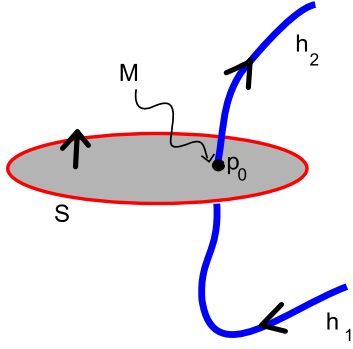, scale=0.8}}
\caption{\label{fi_simple} The simple situation, with two holonomies meeting the surface transversally in one point.}
\end{figure}

Now we will have to specify the action of the functional derivatives more precisely. In particular, we would like it to have a well defined action on holonomy loops. To do this, we borrow some techniques from loop quantum gravity. There, surface integrals of the functional derivative with respect to the connection have been rigorously defined. More precisely, there is an operator  
\begin{equation}
X^{(S)}_{I}\widehat{=}\int_S\epsilon_{cab}\frac{\delta}{\delta A^I_c}\text{d}x^a\wedge \text{d} x^b
\end{equation}
which yields a derivation on holonomy functionals \cite{Ashtekar:1996eg}. Te action of $X^{(S)}$ depends on the orientation of the surface $S$. With $\overline{S}$ denoting $S$ with opposite orientation, one has  
\begin{equation}
X^{(\overline{S})}=- X^{(S)}
\end{equation}
The operator $X^{(S)}_{I}$ can be written as an integral of an operator valued two-form 
${X}_I^{(S)}(p)$ over the surface $S$. This operator valued two-form acts on holonomy loops only when $p$ coincides with a transversal intersection of the surface $S$ with the loop. We will only need the action of this operator in the following simple situation (see figure \ref{fi_simple}): Consider the matrix elements $(h_1[A]Mh_2[A])_{mn}$ of the product of two holonomies $h_1,h_2$ and a constant matrix $M$. These are functionals of the connection. The endpoint $p_0$ of $h_1$ shall be the starting point of $h_2$, and at the same time a transversal intersection point with the surface $S$. Relative orientation of the holonomies and $S$ are assumed to be as indicated in \ref{fi_simple}. In this case, the action of the operator valued two-form is
\begin{equation}
\label{eq:opform}
 {X}^{(S)}_I(p) h[A]Mh'[A]=\frac{1}{2}h[A][M,T_I]_+h'[A] \delta^{(2)}(p,p_0).
\end{equation}
The object $\delta^{(2)}(\cdot,p_0)$ is the two-dimensional delta function on $S$ centered at $p_0$, and $[\cdot,\cdot]_+$ denotes the anti-commutator. We note that ${X}^{(S)}_I(p)$ changes under change of orientation of $S$ by a factor of $-1$,
\begin{equation}
X^{(-S)}=-X^{(S)}.
\end{equation}  

The fact that ${X}^{(S)}_I(p)$ only acts on transversal intersections is crucial. Otherwise, insertion of it into the path-ordered exponential \eqref{eq:nastokes}, with its complicated dependence on holonomies, would almost certainly result in an ill defined operator. 

But when replacing the components of curvature by the operators ${X}^{(S)}_I(p)$ in \eqref{eq:opform}, there remain some issues that we have to resolve:
\begin{itemize}
\item[(a)]  As one can see from \eqref{eq:opform}, the action of ${X}(p)$ is singular, given by a delta function on the surface.
The ordered integration in \eqref{eq:nastokes} will have the consequence that boundaries of integration regions will fall on the support of these delta functions, so we will have to specify what the value of such integrals will be.

\item[(b)] Also evident from \eqref{eq:opform} is the fact that the operators  $X^{(S)}_I(p)$ do not in general commute. In our context, this represents an ordering problem, since the functional derivatives are meant to replace components of curvature, which are certainly commuting. 
\end{itemize}
Let us deal with problem (a) first. Let us imagine that we insert $X^{(S)}_I(p)$ into \eqref{eq:nastokes}, and let the resulting operator act in the simple situation of figure \ref{fi_simple}. In the $n$'th order term, there will be $n$ delta functions centered at the point $p_0$, integrated over in a surface ordered way. Just extracting the relevant integrations, and pulling back into the coordinate space that was used in the statement of the non-Abelian Stokes theorem, we have the structure 
\begin{equation}
\label{eq_sint}
\int_0^1\text{d}s_1\int_0^{s_1}\text{d}s_2\ldots\int_0^{s_{n-1}}\text{d}s_n f(s_1,\ldots s_n)\delta(s_{n},s_0)\delta(s_{n-1},s_0)\ldots \delta(s_{1},s_0).
\end{equation}
We will define the evaluation of this by regularizing the delta functions. 
To this end, let $O((s_1,s_2,\ldots s_n))\equiv O(\boldsymbol{s})$ stand for the ordering of the $n$-tupel
$\boldsymbol{s}=(s_1,s_2,\ldots s_n)$, i.e.\ $O(\boldsymbol{s})$ is a permutation of $\boldsymbol{s}$ such that the elements are now ordered
according to size, increasing from left to right. Then we can rewrite \eqref{eq_sint} as
\begin{equation}
\int_{[0,1]^{n}} \text{d}s_1 \text{d}s_2 \ldots \text{d}s_n\,\,\frac{1}{n!}\delta^{(n)}(\boldsymbol{s_0}, O(\boldsymbol{s}))f(\boldsymbol{s})
\end{equation}
where $\boldsymbol{s_0}=(s_0,s_0,\ldots s_0)$.
We now regularize the delta function in a symmetric fashion, for example
\begin{equation}
\begin{split}
\int_{[0,1]^{n}} \text{d}s_1 \text{d}s_2 &\ldots \text{d}s_n\,\, \frac{1}{n!}\delta^{(n)}(\boldsymbol{s_0}, O(\boldsymbol{s}))f(\boldsymbol{s})\\
&:= \lim_{\epsilon\rightarrow 0}\int_{[0,1]^{n}} \text{d}s_1 \text{d}s_2 \ldots \text{d}s_n\,\,
\frac{1}{n!}\left(\frac{1}{\sqrt{\pi}\epsilon}\right)^n f(\boldsymbol{s}) \exp\left(-\frac{(\boldsymbol{s_0}-O(\boldsymbol{s}))^2}{\epsilon^2}\right)\\
&=\lim_{\epsilon\rightarrow 0}\int_{[0,1]^{n}} \text{d}s_1 \text{d}s_2 \ldots \text{d}s_n\,\,
\frac{1}{n!}\left(\frac{1}{\sqrt{\pi}\epsilon}\right)^n f(\boldsymbol{s}) \exp\left(-\frac{(\boldsymbol{s_0}-\boldsymbol{s})^2}{\epsilon^2}\right)\\
&=\frac{1}{n!} f(\boldsymbol{s_0}).
\end{split}
\end{equation}
Thus, in the evaluation of the $n$ delta-functions, we obtain a factor of $1/n!$ due to the path ordering. 

Now we will start to discuss problem (b), the ordering ambiguity. As the description shows, the action of the operators $X^{(S)}_I(p)$ has two separate aspects: One is what one could call the \emph{external} one, is related to the surface and the intersection with the holonomy loop. The operator only acts on transversal intersections, and it acts locally, in the sense that only the coupling structure at the intersection vertex is changed. These external aspects will not concern us, as they are not directly responsible for the non-commutativity. The other aspect of the action is the \emph{internal} one. 
The operator acts by insertion of elements of the Lie algebra $\lieg$. There is some symmetrization present in the action of $X$, as the insertion is to the left \emph{and} to the right of the matrix $M$. When more than two operators $X$ act at the same point, they do not commute,  nevertheless. In the situation considered here, arbitrarily high powers of the operators $X$ will be needed: Replacement of the curvature $F$ in \eqref{eq:nastokes} by the operators $X$ will give 
\begin{equation}
h_{\partial S}= \um
+\frac{2\pi}{ick}\int_S \text{d}p\, \mathcal{X}(p)
+\left(\frac{2\pi}{ick}\right)^2\iint_{\substack{S\times S\\p_1>p_2}} \text{d}p_1\text{d}p_2\, \mathcal{X}(p_2)\mathcal{X}(p_1)\\
+\ldots
\end{equation}
where the $\mathcal{X}(p)=h_{w(p)} X^{(S)}_I(p)T^I h_{w(p)}^{-1}$, and the path ordering is indicated by the notation $p_1>p_2$. The integrand of the n'th order term thus reads
\begin{equation}
h_{w(p_n)} X^{(S)}_{I_n}(p_n)T^{I_n} h_{w(p_{n})}^{-1}\,
h_{w(p_{n-1})} X^{(S)}_{I_{n-1}}(p_{n-1})T^{I_{n-1}} h_{w(p_{n-1})}^{-1} 
\ldots 
h_{w(p_1)} X^{(S)}_{I_1}(p_1)T^{I_1} h_{w(p_1)}^{-1}.
\end{equation}
The $X^{(S)}_I$ commute unless they are evaluated at the same point. From what we have said before, it should however be clear that they will have to be evaluated at the same point, as they are concentrated on intersections of $S$ with holonomy loops. At an intersection point $p$, the above thus becomes 
\begin{equation}
X^{(S)}_{I_n}(p)X^{(S)}_{I_{n-1}}(p)\ldots X^{(S)}_{I_1}(p)\,
h_{w(p)}T^{I_{n}}T^{I_{n-1}}\ldots T^{I_1}h_{w(p)}^{-1}. 
\label{eq:toorder}
\end{equation}
The crucial point is that the ordering for the $X^{(S)}_{I}(p)$ chosen in the above expression is by no means the only possible one, as the $X$ replace the components of $F$, which are commuting. It is not clear a priori, which ordering needs to be chosen to reproduce the known expectation values for holonomy loops in CS theory (see appendix \ref{ap_jones}). In fact, we first considered more or less natural choices, such as fully symmetric ordering, or an ordering derived from the ordering present in the integration paths of \eqref{eq:nastokes}, but without success. It turns out, however, that there is a mathematically preferred ordering, which will do the job. We will describe it in detail in the following section. 
\section{The use of the Duflo isomorphism}
\label{se_duflo}
As we have seen in the preceding section, the operators  $X^{(S)}_I(p)$ do not commute when acting at the same point, and we will have to find a suitable ordering. 
To this end, we will show that the problem can be considered as that of quantization of a certain space of functions. This will lead us to introduce an ordering that is mathematically preferred. 

To start out, note that under gauge transformations, the curvature components $F_I$ that we want to represent by operators $X^{(S)}_I$ transform like the components of an element of $\lieg$. But the $F_I$ commute, thus it is not too useful to view them as elements of $\lieg$. Rather, we are led to think of them as linear functions on the dual space $\lieg^*$. The $F_I$ are also two-forms over the tangent space of $M$, but this does not concern us here -- it is reflected in what we had called the \emph{external} action of the $X$ and has nothing to do with the ordering problem. Thus, let us introduce quantities $E_I$ that are \emph{just} linear functions on the dual space $\lieg^*$, and let us phrase the discussion in terms of them. (One can think of $E_I$ as $F_{Iab}s^{ab}$ for some anti-symmetric $s^{ab}$.) Functions can be added and multiplied pointwise, and in this way the $E_I$ generate the algebra of polynomial functions over $\lieg^*$. This algebra is also called the \emph{symmetric algebra} over $\lieg$, and is denoted by $\sym(\lieg)$.\footnote{An alternative definition for $\sym(\lieg)$ is as follows: Take the free algebra (also called tensor algebra) over the \emph{vector space} $\lieg$ and divide by the ideal generated by the elements $v\otimes v'-v'\otimes v$, where v,v' are in $\lieg$.} It is obviously a commutative algebra.  
 
The internal action of the operators  $X^{(S)}_I$ is by invariant vector fields which, as differential operators on functions on $G$, form the \emph{universal enveloping algebra} $U(\lieg)$.\footnote{An alternative definition of $U(\lieg)$ is as follows: Take the free algebra (also called tensor algebra) over the \emph{vector space} $\lieg$ and divide by the ideal generated by the elements $v\otimes v'-v'\otimes v -[v,v']$, where v,v' are in $\lieg$.} This algebra is non-commutative. The ordering problem arises, because the commutative algebra $\sym(\lieg)$ must be mapped to  
the non-commutative algebra $U(\lieg)$. What is the relevant class of maps? 
$\sym(\lieg)$ carries a natural Poisson bracket given by 
\begin{equation}
\{E_I,E_J\}=f_{IJ}{}^KE_K,
\label{eq:poi}
\end{equation}
and extension to more complicated functions by linearity and Leibniz rule. $f_{IJ}{}^K$ denotes the structure constants of $\lieg$. In the light of the above Poisson structure, the replacement of $E_I$ by the operator 
\eqref{eq:opform} can be viewed as (a suitably symmetrized version of) the standard quantization map  
\begin{equation}
E_I\mapsto X_I 
\label{eq:qmap}
\end{equation}
from the linear part of $\sym(\lieg)$ to $U(\lieg)$. Here we have used the notation $X_I$ for the (left-)invariant vector fields, viewed as generators of $U(\lieg)$.
What we are seeking for the ordering problem is an extension of the map \eqref{eq:qmap} to more general polynomials, i.e.\ a \emph{quantization} of the Poisson structure \eqref{eq:poi}. As is well known -- compare for example the Groenewald-van Hove Theorem\footnote{See \cite{gs} for the original works, and \cite{am} for a more recent exposition.} in the case of a trivial phase space -- there is no extension that will map Poisson brackets to commutations for \emph{all} functions in $\sym(\lieg)$. A natural quantization map $\chi:\sym(\lieg)\rightarrow U(\lieg)$ which is suitable for many purposes, is the so-called symmetric quantization. It can be defined by its action  
\begin{equation}
\chi:\, E_{I_1}E_{I_2}\ldots E_{I_n} \longmapsto \frac{1}{n!}\sum_{\pi\in S_n} X_{I_{\pi(1)}}X_{I_{\pi(2)}}\ldots X_{I_{\pi(n)}}. 
\end{equation}
on monomials. We will shortly see, however, that for our purposes it is crucial to improve on $\chi$. To get there, we note that we are especially interested in mapping elements of $\sym(\lieg)$ that are invariant under the action of $G$. The space of such elements is commonly denoted $\sym(\lieg)^\lieg$. Consider again the n'th order integrand evaluated at a point $p$, \eqref{eq:toorder},  from the last section. Taking the trace, and using the notation from this section, it corresponds to the element
\begin{equation}
\tr( E^n)\equiv \tr(\underbrace{E\cdot E\cdot\ldots\cdot E}_{n \text{ times} }) 
\end{equation}
of $\sym(\lieg)$, 
where we have used $E:=E_I T^I$, and we remind the reader that $T^I=k^{IJ}T_J$, i.e.\ some operator in the defining representation, not an element of the dual. This expression is invariant under the action of $G$.
Thus the space $\sym(\lieg)^\lieg$ merits special attention in the ordering problem. 

Coming back to the discussion of quantization maps, it should be expected, that `good' quantization maps take 
$\sym(\lieg)^\lieg$ to the sub-algebra $U(\lieg)^\lieg$ of invariant differential operators (``Casimir operators'') under the action of $G$. We also note that it is easily seen that $U(\lieg)^\lieg$ is the center of $U(\lieg)$ and hence commutative.  Looking at symmetric quantization, we see that $\chi$ maps $\sym(\lieg)^\lieg$ to $U(\lieg)^\lieg$, but it is not a homomorphism of algebras. But $\chi$ can be deformed in such a way that it becomes an isomorphism between those sub-algebras. That means  
there is a quantization map that is optimal for $\sym(\lieg)^\lieg$, in the sense that it restricts to an isomorphism between  $\sym(\lieg)^\lieg$ and $U(\lieg)^\lieg$. This quantization map is the \emph{Duflo isomorphism} (Harish-Chandra, in full generality: Duflo \cite{duflo}), and we will use it. We describe it in detail, now.  
   
We define the following function on $\lieg$:
\begin{equation}
j^\frac{1}{2}(x)
=\det{}^\frac{1}{2}\left(\frac{\sinh\frac{1}{2}\ad x}{\frac{1}{2}\ad x}\right).
\label{eq:j12}
\end{equation}
$\ad x$ is the linear map given by the adjoint action of $\lieg$ on itself, and functions of it are defined either as power series, or through functional calculus.  We note that the above map is invariant under change of basis, as it is given by the determinant of a linear map. 
Later, it will also be very convenient to know a series expansion in terms of traces. Using that $\det\exp=\exp\tr$, one finds 
\begin{equation}
j^\frac{1}{2}(x)
=\exp\left(\sum_{n\geq 1}\frac{b_{2n}}{4n(2n)!}\tr[(\ad x)^{2n}]\right)
= 1 +\frac{1}{48} \norm{x}^2 +\ldots,
\end{equation}
where $b_n$ are the Bernoulli numbers. 
$j^\frac{1}{2}(x)$ can be used to define a differential  operator $j^\frac{1}{2}(\partial)$ on $\sym(\lieg)$, by substituting for the components of $x$ derivatives with respect to the generators of $\sym(\lieg)$. To be very explicit, fix some basis $\{T_I\}_I$ of $\lieg$. Then $x=x^IT_I$ and $j^\frac{1}{2}$ is a function of the components $x^I$. There are natural derivatives on $\sym(\lieg)$,
\begin{equation}
\partial^I\equiv \frac{\partial}{\partial E_I}, \qquad \partial^I E_J=\delta_J^I. 
\end{equation} 
Now substitute for the variables $x^I$ in $j^\frac{1}{2}(x)$ the derivatives $\partial^I$ to obtain $j^\frac{1}{2}(\partial)$. 

The Duflo isomorphism is now the map 
\begin{equation}
\Upsilon = \chi\circ j^{\frac{1}{2}}(\partial).
\label{eq:duflo}
\end{equation}
It is indeed an isomorphism of the invariant sub-algebras, but this is not easy to prove.  
We will soon see that, at least in the simple cases that we will consider, the terms appearing in the non-abelian Stokes formula can indeed be interpreted as elements of $\sym(\lieg)^\lieg$, and thus we can use the Duflo isomorphism to solve the ordering problem for us. Surprisingly, we will find that this gives the right result. 

Let us finish this section by pointing out that the situation in loop quantum gravity, from which we had borrowed the definition \eqref{eq:opform} for the functional derivative that was the starting point of our discussion of the ordering problem, fits quite nicely into the abstract framework that we have laid out above:
There, the internal action of $X^{(S)}_I(p)$ is also the result of the quantization of the components of an element $E_I(x)t^I$ in the \emph{dual} of $\lieg$ by invariant vector fields, and higher powers of $X$ represent the quantization of more general functions on the algebra dual $\lieg^*$.
The meaning of the Poisson structure \eqref{eq:poi} in LQG is not entirely obvious, but it can be argued that it is simply a necessary consequence of the Poisson structure between the field $E(x)$ and the configuration variables (see \cite{Sahlmann:2010zf}). For different takes on the non-commutativity of the $X^{(S)}_I(p)$ see also \cite{Ashtekar:1998ak} and \cite{Baratin:2010nn}.

\section{Unknots}
\label{se_unknots}
In this section, we will discuss the calculation of the expectation value of unknots in SU(2) and SU(3) CS theory. Let $\alpha$ be a trivial loop in $S^3$, and $S$ be a disc with boundary $\partial S=\alpha$. Then we have the operator
\begin{equation}
\boldsymbol{W_S}=\tr\left[\ppexp{S} \left(\frac{2\pi}{ick}\mathcal{X}^{(S)}\right)\right]
\end{equation}
where 
\begin{equation}
\mathcal{X}^{(S)}(p)=h_{w(p)}{X}^{(S)}(p)h_{w(p)}^{-1}
\end{equation}
is the operator corresponding to the functional derivative, parallel transported to the origin of the surface. We thus have (formally)
\begin{equation}
\begin{split}
\expec{W_\alpha}&=\int_{\abar} W_\alpha[A] \exp(i
S_{\text{CS}}[A])\, \mal [A]\\
&=\int_{\abar}\boldsymbol{W_S}[\exp(i S_{\text{CS}}[A])]\, \mal [A]\\
&=\int_{\abar}\boldsymbol{W_S}^\dagger[\one]\exp(i S_{\text{CS}}[A])\, \mal [A]
\end{split}
\end{equation}
with 
\begin{equation}
\boldsymbol{W_S}^\dagger=\tr\left[\ppexp{S} \left(\frac{2\pi i}{ick}\mathcal{X}^{(S)}\right)\right].
\label{eq:wdagger}
\end{equation}
Thus we need to calculate  $\boldsymbol{W_S}^\dagger[1]$. 
If we were to use $\boldsymbol{W_S}^\dagger$ without further ordering, we would just obtain $\tr(\um)$, where $\um$ is the unit matrix in the defining representation, giving 2 or 3, for SU(2) and SU(3), respectively. This is the case because $X^{(S)}_I(p)[1]=0$. We will however order the internal action of $X$ using the Duflo map. Then, if more than one $X$ acts at the same point, there will be a non-trivial ordering which, as we will see, leads to a non-trivial result for  $\boldsymbol{W_S}^\dagger[1]$.
Thus, let us consider the integrand in \eqref{eq:wdagger}, at a certain order $n$ of the exponential, 
\begin{equation}
\left(\frac{2\pi i}{ick}\right)^n \tr\left[X^{(S)}(p_1)h_{w(p_1,p_2)}X^{(S)}(p_2)h_{w(p_2,p_3)}\ldots X^{(S)}(p_n)h_{w(p_n,p_1)}\right],
\label{eq:norder}
\end{equation}
where $w(p,p')=w(p')\circ w(p)^{-1}$. The points are taken to be ordered. The action of this term on $\one$ is zero as long as the points are not coinciding, but may be non-zero for two or more coinciding points. 

We will now discuss the case, where \emph{all} points coincide. We will drop the terms where there are two or more sets of coinciding points.
We can not fully justify this for the case of trivial loops. We will see that this happens automatically for the case where there are intersections between holonomy loops and the surface, so there is no need for this ad hoc choice in other cases.   
Thus, let $p$ be some point in $S$, and consider $p_1=p_2=\ldots=p_n=p$ in \eqref{eq:norder}. All the holonomies are then just the identity. Taking all orders together, we thus consider the quantization of the following function on $\lieg^*$: 
\begin{equation}
W_p=\tr \exp\left( \frac{2\pi i}{ck}E\right)=\sum_{n\geq 0}\frac{1}{n!}\left( \frac{2\pi i}{ck}\right)^n \tr(E^n). 
\label{eq:w}
\end{equation}
Here, as in the preceding sections, the $E_I$ are linear functions on $\lieg^*$, and we write $E=E_I T^I$, which is a matrix of linear functions. 
We will now quantize the above function using the Duflo map. (See the discussion in the previous section for motivation and further explanation.)
Closer inspection reveals that $E$ enters just in terms of $\betr{E}^2:=\tr(E^2)$ (or, equivalently $\norm{E}^2:=\betr{E}^2/c$) for both groups, and additionally on  
\begin{equation}
D(E):=\tr(E^3)
\end{equation}
for SU(3). In fact, some well known matrix identities can be used to calculate the traces in \eqref{eq:w} recursively. From the Cayley-Hamilton Theorem (``a matrix satisfies its own characteristic equation'') and the relation between $\det$ and $\tr$, it follows that 
\begin{equation}
\qquad M^2=\frac{1}{2}\tr(M^2)\um, \qquad  A^3=\frac{1}{2}\tr(A^2)A+\frac{1}{3}\tr(A^3)\um
\end{equation}
for $2\times 2$ matrices $M$ and $3\times 3$ matrices $A$ that are trace-free.  Using this, we find
\begin{equation}
\tr(E^0)=2,\quad \tr(E^1)=0, \qquad \tr(E^s)=\frac{1}{2}|E|^2 \tr(E^{s-2})
\label{eq:rec1}
\end{equation}
for SU(2) and 
\begin{equation}
\begin{split}
\tr(E^0)=3,\quad \tr(E^1)=0, \quad \tr(E^2)= |E|^2,\\ \tr(E^s)=\frac{1}{2}|E|^2\tr(E^{s-2})+\frac{2}{3}D\tr(E^{s-3})
\end{split}
\label{eq:rec2}
\end{equation}
for SU(3). Since $\Upsilon$ is an algebra-homomorphism, we do not need to calculate $\Upsilon(W_P(E))$ in one stroke, but can instead calculate $\Upsilon(\norm{E}^2)$, $\Upsilon(D(E))$ and insert the results into \eqref{eq:w}. This simplifies the calculation drastically, because $\norm{E}^2, D(E)$ are only second and third order, respectively, in $E$, and thus to calculate their image under $\Upsilon$, we only need terms up to two derivatives in $j^{1/2}(\partial)$ -- the higher order terms will just annihilate the functionals. Thus we calculate 
\begin{equation}
j^\frac{1}{2}(\partial)\norm{E}^2
=\norm{E}^2+\frac{2}{48}k^{IJ}k_{IJ}
=\norm{E}^2+
\begin{cases}
\frac{1}{8} &\text{ SU(2)}\\
\frac{1}{3} &\text{ SU(3)}
\end{cases}
\end{equation}
In the case $G=SU(3)$, one has to deal with the additional function $D(E)$: 
The second order term in $j^{1/2}(\partial)$ vanishes when applied to $D(E)$ due to the fact that $\tr(T_IT_JT_K)k^{IJ}=0$. There is no third order term, and the fourth and higher order terms vanish again, since they have more than three derivatives. Thus 
\begin{equation}
j^\frac{1}{2}(\partial) D(E)=D(E)
\end{equation}
Altogether we find 
\begin{equation}
\Upsilon(\norm{E}^2)=\Delta +\text{id}\begin{cases}
\frac{1}{8} &\text{ SU(2)}\\
\frac{1}{3} &\text{ SU(3)}
\end{cases},\qquad \text{with } \Delta=k^{IJ}X_{I}X_{J}
\end{equation}
where $\{X_I\}_I$ is a basis of $U(\lieg)$ (the left invariant vector fields, say) and 
\begin{equation}
\Upsilon(D(E))=\tr(T^{(I}T^JT^{K)})X_IX_JX_K. 
\end{equation}
The symmetrization on the right hand side in the last formula comes from the action of the symmetrization map $\chi$, and is non-trivial.  
With these results on hand, we can now calculate $\boldsymbol{W_S}^\dagger[1]$, with the understanding that the Duflo map has been used to order it, and that all terms but the ones in which all action points coincide have been dropped. The elements of $U(\lieg)$ act as left-invariant derivatives. For notational convenience, we will not distinguish this action from the elements themselves. We note that $\Delta[1]=0$. For SU(2) we find 
\begin{equation}\begin{split}
\boldsymbol{W_S}^\dagger[1]&=C\Upsilon \left[\tr \exp\left( -\frac{2\pi i}{ck}E\right)\right][1]\\
&=2C+2C\sum_{n > 0}\frac{1}{(2n)!}\frac{1}{2^n}\left( \frac{2\pi i}{ck}\right)^{2n} \Upsilon(|E|^{2n})[1]\\
&=2C+2C\sum_{n > 0}\frac{1}{(2n)!}\left(\frac{c}{2}\right)^n\left( \frac{2\pi i}{ck}\right)^{2n} (\Delta + \frac{1}{8})^n[1]\\
&=2C+2C\sum_{n > 0}\frac{1}{(2n)!}\left( \frac{2\pi i}{8ck}\right)^{2n}\\
&=2C+2C\sum_{n > 0}\frac{1}{(2n)!}\left( \frac{\pi i}{k}\right)^{2n}\\
&=C(q^\frac{1}{2}+q^{-\frac{1}{2}})\\
&=C\frac{q-q^{-1}}{q^\frac{1}{2}-q^{-\frac{1}{2}}}.
\end{split}\end{equation}
In the above calculation, $C$ stands for an (infinite) constant that comes from the fact that \emph{each} point in the surface gives a contribution \eqref{eq:w} to the surface ordered exponential. There is a freedom in CS theory to chose a normalization, and we will chose this to set $C=1$. 
Thus one finds 
\begin{equation}
\expec{W_\alpha}=\frac{q-q^{-1}}{q^\frac{1}{2}-q^{-\frac{1}{2}}}
\end{equation}
We note that our argument can be carried out in exactly the same way for a collection $\alpha_1,\alpha_2,\ldots\alpha_n$ of unknotted, unlinked loops, giving
\begin{equation}
\expec{W_{\alpha_1}W_{\alpha_2}\ldots W_{\alpha_n}}=\left[\frac{q-q^{-1}}{q^\frac{1}{2}-q^{-\frac{1}{2}}}\right]^n
\end{equation}
One can also carry out the calculation for $G=$SU(3). 
Using the fact that $\Upsilon(D(E))[1]=0$, the recursion 
\eqref{eq:rec2} simplifies drastically, and one finds 
\begin{equation}\begin{split}
\boldsymbol{W_S}^\dagger[1]&=\Upsilon\tr \exp\left( -\frac{2\pi i}{ck}E\right)\\
&=3+2\sum_{n\geq 0}\frac{1}{(2n)!}\frac{1}{2^n}\left( \frac{2\pi i}{ck}\right)^{2n} \Upsilon(|E|^{2n})[1]\\
&=3+2\sum_{n\geq 0}\frac{1}{(2n)!}\left(\frac{c}{2}\right)^n\left( \frac{2\pi i}{ck}\right)^{2n} (\Delta + \frac{1}{3})^n[1]\\
&=3+2\sum_{n\geq 0}\frac{1}{(2n)!}\left( \frac{2\pi i}{6ck}\right)^{2n}\\
&=3+2\sum_{n\geq 0}\frac{1}{(2n)!}\left( \frac{2\pi i}{k}\right)^{2n}\\
&=1+q+q^{-1}\\
&=\frac{q^{\frac{3}{2}}-q^{-\frac{3}{2}}}{q^\frac{1}{2}-q^{-\frac{1}{2}}}
\end{split}\end{equation}
giving 
\begin{equation}
\expec{W_{\alpha_1}W_{\alpha_2}\ldots W_{\alpha_n}}=\left[\frac{q^{\frac{3}{2}}-q^{-\frac{3}{2}}}{q^\frac{1}{2}-q^{-\frac{1}{2}}}\right]^n
\end{equation}
for a system $\alpha_1,\alpha_2,\ldots\alpha_n$ of unknotted, unlinked loops.
\section{Hopf link}
\label{se_hopf}
In this section, we will discuss the first non-trivial link, the \emph{Hopf link}, consisting of two simply linked, but unknotted loops $\alpha_1$ and $\alpha_2$. In fact, when taking into account orientation, there are two different Hopf links, see figure \ref{fi_hopfpm}. Let us also fix an orientation of $S^3$. We will first consider the Hopf link $H_+$.
\begin{figure}
\centerline{\epsfig{file=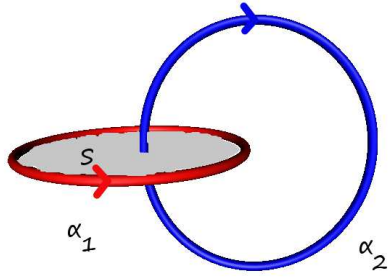, scale=0.27}}
\caption{\label{fi_hopf_link2} The Hopf link $H_+$, already drawn with surface $S$}
\end{figure}
We pick a disc $S$ that has $\alpha_1$ as a boundary, and one intersection with $\alpha_2$, see figure \ref{fi_hopf_link2}. We will chose the orientation of $S$ consistent with that of $\alpha_1$ and the chosen orientation on $S^3$. Again one proceeds as per the strategy outlined in the introduction: 
\begin{equation}\begin{split}
\expec{W_{\alpha_2}W_{\alpha_1}}&=\int_{\abar} W_{\alpha_2}W_{\alpha_1}[A] \exp(i
S_{\text{CS}}[A])\, \mal [A]\\
&=\int_{\abar}W_{\alpha_2}\boldsymbol{W_S}[\exp(i S_{\text{CS}}[A])]\, \mal [A]\\
&=\int_{\abar}\boldsymbol{W_S}^\dagger[W_{\alpha_2}]\exp(i S_{\text{CS}}[A])\, \mal [A]
\end{split}\end{equation}
Thus the object of interest is $\boldsymbol{W_S}^\dagger[W_{\alpha_2}]$. 
Now the action of all the operators $X^{(S)}$ is concentrated at the point $p_0$ of intersection between $S$ and $\alpha_2$ (see \eqref{eq:opform}), so there is no factor $C$ and no need to drop any terms by hand. The elements of $U(\lieg)$ now act in the defining representation. In particular, for SU(2) one has 
\begin{equation}\begin{split}
\Upsilon(\norm{E}^2)[W_{\alpha_2}]
&=\Upsilon(\norm{E}^2)[\tr(h_{\alpha_2})]\\
&=\left(\Delta+\frac{1}{8}\right)[\tr(h_{\alpha_2})]\\
&=\tr\left(\left(k^{IJ}T_IT_J+\frac{1}{8}\um\right)h_{\alpha_2}\right)\\
&= \tr\left(\frac{1}{2}\um h_{\alpha_2}\right)\\
&= \frac{1}{2}W_{\alpha_2}
\end{split}\end{equation}
because $k^{IJ}T_IT_J=3/8\um$ in the defining representation (see appendix \ref{ap_lie}). 
This we can apply to obtain
\begin{equation}\begin{split}
\boldsymbol{W_S}^\dagger[W_{\alpha_2}]&=\tr \exp\left( -\frac{2\pi i}{ck}E\right)\\
&=2+2\sum_{n > 0}\frac{1}{(2n)!}\frac{1}{2^n}\left( \frac{2\pi i}{ck}\right)^{2n} \Upsilon(|E|^{2n})[W_{\alpha_2}]\\
&=\left[2+2\sum_{n > 0}\frac{1}{(2n)!}\left(\frac{c}{2}\right)^n\left( \frac{2\pi i}{ck}\right)^{2n} \left(\frac{1}{2}\right)^n\right]W_{\alpha_2} \\
&=\left[2+2\sum_{n > 0}\frac{1}{(2n)!}\left(\frac{1}{16}\right)^n\left( \frac{2\pi i}{ck}\right)^{2n}\right]W_{\alpha_2}\\
&=\left[2+2\sum_{n > 0}\frac{1}{(2n)!}\left( \frac{2\pi i}{k}\right)^{2n}\right]W_{\alpha_2}\\
&=(q+q^{-1})W_{\alpha_2}.
\end{split}\end{equation}
Thus so far one has 
\begin{equation}
\expec{W_{\alpha_2}W_{\alpha_1}}=(q+q^{-1})\expec{W_{\alpha_2}}.
\end{equation}
But the latter expectation value has been computed above, so 
\begin{equation}
\expec{W_{\alpha_2}W_{\alpha_1}}=(q+q^{-1})(q^{\frac{1}{2}}+q^{-\frac{1}{2}})=q^{\frac{3}{2}}+q^{\frac{1}{2}}+q^{-\frac{1}{2}}+q^{-\frac{3}{2}}=q^{-\frac{3}{2}}\left(q^3+q^2+q^1+1\right).
\end{equation}
Thus the result obtained is, up to a power of $q^{1/2}$, the well known result. But the latter is obtained with standard framing (see appendix \ref{ap_jones}), and calculations in other framings differ by powers of $q^{1/2}$. Thus our approach gives the result here with respect to a framing that differs from standard framing. In fact, the factor $q^{3/2}$ would be expected \emph{precisely} for a change from standard framing to the trivial framing associated with the surfaces chosen for the components $\alpha_1$ and $\alpha_2$, hence the result is very transparent.

We can also do the calculation for the mirror Hopf-link $H_-$. It can be obtained from either changing the orientation of $\alpha_1$ or that of $\alpha_2$. In either case, the net result is that there is a factor of $-1$ in the action of $X^{(S)}$ on $W_{\alpha_2}$: If one changes orientation of $\alpha_1$, orientation of $S$ will change accordingly, and hence $X^{(S)}$ will acquire a minus sign. If on the other hand the orientation of $\alpha_2$ is
changed, the situation differs from the standard setup depicted in figure \ref{fi_simple} again by the reversal of orientation of the surface, and hence a sign. In any case, since only even powers of $X^{(S)}$ act, the change of sign is irrelevant, and the result of the entire calculation will be unchanged.
Equivalently, one sees that changing the orientation of either $\alpha_1$ or $\alpha_2$ will effectively replace $q$ by $q^{-1}$ in the entire calculation, giving  
\begin{equation}
\expec{H_-}=q^{\frac{3}{2}}\left(q^{-3}+q^{-2}+q^{-1}+1\right)
\end{equation}
This is consistent with the fact that the Jones polynomials of the two Hopf-links are related by multiplication with a power of $q^{1/2}$. Here, however, the framing in which the above is valid is neither standard framing, nor the trivial framing that one would expect. It would be very good to understand the origin of this framing better. 

Now we come to the calculation for SU(3). The situation is more complicated here, since the quantization of $D(E)$ acts in a non-trivial way, unlike in the case of the unknot. Let us calculate this action first. Note that, as argued in the case of the unknot, $j^{1/2}(\partial)D(E)=D(E)$. The symmetrization $\chi$ is non-trivial in this case, so that we obtain  
\begin{equation}
\Upsilon(D(E))=T^{(I}T^{J}T^{K)}\tr(T_IT_JT_K).
\end{equation}
Thus 
\begin{equation}\begin{split}
\Upsilon(D(E))[W_{\alpha_2}]
&=\tr(T_IT_JT_K)T^{(I}T^{J}T^{K)}[\tr(h_{\alpha_2})]\\
&=\tr(T_IT_JT_K)\tr\left(T^{(I}T^{J}T^{K)}h_{\alpha_2}\right)\\
&= \tr\left(  \frac{5}{486}\um h_{\alpha_2}\right)\\
&= \frac{5}{486}W_{\alpha_2}. 
\label{eq:ac1}
\end{split}\end{equation}
See appendix \ref{ap_lie} for a calculation of the traces.
Similarily, we find  
\begin{equation}\begin{split}
\Upsilon(\norm{E}^2)[W_{\alpha_2}]
&=\Upsilon(\norm{E}^2)[\tr(h_{\alpha_2})]\\
&=\left(\Delta+\frac{1}{3}\right)[\tr(h_{\alpha_2})]\\
&=\tr\left(\left(k^{IJ}T_IT_J+\frac{1}{3}\um\right)h_{\alpha_2}\right)\\
&= \tr\left(\left(\frac{4}{9}+\frac{1}{3}\right)\um h_{\alpha_2}\right)\\
&= \frac{7}{9}W_{\alpha_2}.
\label{eq:ac2}
\end{split}\end{equation}
Thus $W_{\alpha_2}$ will be an eigenfunction of $\boldsymbol{W_S}^\dagger$, with an eigenvalue $c_{S,\alpha_2}$ that we will determine shortly. We thus obtain
\begin{equation}
\expec{H_+}
=\int_{\abar}\boldsymbol{W_S}^\dagger[W_{\alpha_2}][A]
\exp(i S_{\text{CS}}[A])\, \mal [A]
=c_{S,\alpha_2}\int_{\abar}W_{\alpha_2}[A]
=c_{S,\alpha_2}(1+q+q^{-1})
\end{equation} 
Comparing with the SU(3) Jones polynomial of $H_+$, we expect 
\begin{equation}
c_{S,\alpha_2}=q^\delta(q^{-1}+q^{-3}+q^{-4})=:f_\delta,
\end{equation}
where $q^{\delta}$ is due to possible framing different than standard framing. To check this, it is easiest to view both $c_{S,\alpha_2}$ and $f_\delta$ as a power series in 
$2\pi i/k$ and compare coefficients. Certainly 
\begin{equation}
f_\delta=\sum_{n=0}^\infty\frac{1}{n!}c_n \left(\frac{2\pi i}{k}\right)^n, \qquad \text{with } c_n=(\delta-1)^n+(\delta-3)^n+(\delta-4)^n. 
\end{equation}
For  
\begin{equation}
c_{S,\alpha_2}=\sum_{n=0}^\infty\frac{1}{n!}c'_n \left(\frac{2\pi i}{k}\right)^n
\end{equation}
we compute the first terms from \eqref{eq:rec2}, \eqref{eq:ac2}
to be 
\begin{equation}
c'_0=3, \qquad c'_1=0\qquad c'_2= \frac{1}{c}\frac{7}{9}=\frac{14}{3}.
\end{equation}
Note that the factor $1/c$ for $c'_2$ is the result of the $c$ in $|E|^2=c\norm{E}^2$ canceling against $1/c^2$ from the replacement  \eqref{eq:func}. $c'_1=0$ can only be correct, if $c_1=0$, which fixes 
$\delta=8/3$. For this choice of $\delta$, one also finds $c_2=14/3=c'_2$. 
Moreover, it is easy to check that the coefficients $c_n$ satisfy the recursion relation  
\begin{equation}
c_n=\frac{7}{3}c_{n-2}+\frac{20}{27}c_{n-3}. 
\end{equation}
But the $c'_n$ satisfies
\begin{equation}
c_n=a c_{n-2}+b c_{n-3},
\end{equation}
where $a$ and $b$ are the eigenvalues of the operators 
\begin{equation}
\frac{1}{2 c^2}\Upsilon(|E|^2), \qquad \frac{1}{3 c^3}\Upsilon(D(E)),
\end{equation}
respectively, acting on $W_{\alpha_2}[A]$. Due to \eqref{eq:ac1}, \eqref{eq:ac2}, one finds $a=7/3$, $b=20/27$, whence 
$c_{S,\alpha_2}=f_\delta$, and 
\begin{equation}
\begin{split}
\expec{H_+}
&=q^\frac{8}{3}(q^{-1}+q^{-3}+q^{-4})(1 +q + q^{-1})\\
&=q^\frac{8}{3}(1+q^{-1}+2q^{-2}+2q^{-3}+2q^{-4}+q^{-5}).
\end{split}
\end{equation}
This is fully consistent with previous work, including the factor $q^{8/3}$ that would \emph{precisely} be expected for a change from standard framing to the trivial framing associated with the surfaces chosen for the components $\alpha_1$ and $\alpha_2$ (see appendix \ref{ap_jones}). 

The calculation can again be done in exactly the same way for the Hopf-link $H_-$, and as was explained before, all that changes is that $q$ gets replaced by $q^{-1}$. Thus one finds 
\begin{equation}
\expec{H_-}
=q^{-\frac{8}{3}}(1+q^{1}+2q^{2}+2q^{3}+2q^4+q^{5}).
\end{equation}
This is fully consistent with previous work (see appendix \ref{ap_jones}), but now the framing is different than one would expect. 

\begin{figure}
\centerline{\epsfig{file=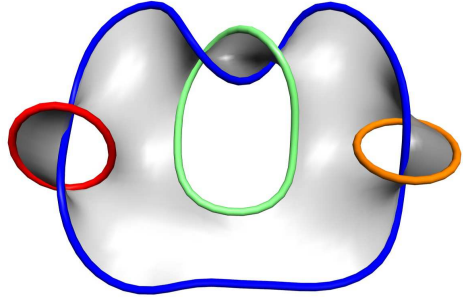, scale=0.15}}
\caption{\label{fi_keychain} An example for the `key chain link': $L_3$ with its Seifert surface}
\end{figure}

In a slight extension of the above results, consider the link that is formed by $n$ unknots $\alpha_1, \ldots \alpha_n$ that are all unlinked with respect to each other, but all linked to one additional unknot $\alpha_0$, like in a key chain. Denote this link with $L_n$ (we do not worry about the orientations here). See also figure \ref{fi_keychain}. The Hopf links considered above, are just a special case, $L_1$. Indeed, since $\alpha_1, \ldots \alpha_n$ are only linked with $\alpha_0$, one can eliminate them one by one by turning them into surface operators and repeating the calculation above. One finds 
\begin{equation}
\expec{L_n}=q^{-\frac{N}{2}}(q^{\frac{1}{2}}-q^{-\frac{1}{2}})\expec{L_{n-1}}+q^{-n}\expec{\bigcirc}\expec{L_{n-1}}
\end{equation} 
with $N=2$ for SU(2), $N=3$ for SU(3), as it should be, up to an overall factor due to non-standard framing. 

\section{Discussion}
\label{se_disc}
In the previous sections, we have calculated expectation values of various holonomy functionals in SU(2) and SU(3) Chern Simons theory. Our results reproduce the known results. This is a remarkable success for the method. 
It shows that the formal manipulations under the path integral are well justified, and that the Duflo isomorphism has a very special status as an ordering map for the functional derivatives. 

At this point, we should also mention some drawbacks of the method. In the calculation of the unknot expectation values, we had to make certain ad hoc choices, in particular of dropping certain terms, and removing an infinite constant. Moreover, the method seems quite tedious to apply in practice. In particular, in other approaches to the calculation of CS expectation values, 
the derivation of the skein relations is the fundamental step. The expectation values then follow. We did not find any useful interpretation of the skein relations in terms of non-infinitesimal surfaces bounding loops, so we did not see a direct way of deriving them in our context. It should however be interesting to carefully revisit the derivation of the Skein relations in \cite{Gambini:1996mb} to see wether it can be adapted to the present setting. One more drawback is that the method does not give any obvious handle on the (nontrivial) value of the path integral without any holonomy loop insertion.  

Certainly, the calculations done in the present work constitute only a finite number of tests of the approach. It should certainly be applied in more general situations. Let us comment on several particular possibilities: 

\textbf{General groups:} The ingredients which were used in the new approach, most notably the  generalized Stokes theorem, the Duflo map, and the functional derivatives, are available for more general groups. In fact,  Stokes theorem and the Duflo map work for general Lie groups. The LQG machinery is available for compact Lie groups, but extensions may be possible, too. Regarding our results, one would need the values of the Casimir elements in the defining representations, and the breakdown of the more complicated invariants into Casimirs, in analogy to \eqref{eq:rec1}, \eqref{eq:rec2}, which should be obtainable with limited amounts of group theory, at least for SU(N).

\textbf{General representations, spin networks:} One could also consider expectation values in representations other that the defining representation of the structure group. Again, moderate amounts of group theory should be sufficient to achieve this. Moreover, the techniques that we used seem in principle sufficient to also compute expectation values for \emph{spin networks} i.e.\ gauge invariant functions of holonomies along edges of a graph embedded in the manifold.  

\textbf{Genuine knots:} Up to now, we have only considered loops that were not knotted (although they were linked). Can the new approach also handle non-trivial knots? Consider one such knot. The first idea would be to use a Seifert surface in the non-abelian Stokes theorem. Since the Seifert surface  would have no self-intersections, one would obtain an operator that is fairly trivial, and obtain the same expectation value as for the unknot. But this reasoning is flawed, due to the fact that the non-abelian Stokes theorem is applicable only to surfaces that are simply connected. One would thus have to use different surfaces, in general with self-intersections. Those self-intersections would render the action of the operator quite non-trivial, but they would in fact risk making it ill defined: There are families of holonomies routed inside of the surface, and they may intersect the surface transversally near the points of self-intersection. One has some freedom in the routing of those holonomies in the non-abelian Stokes theorem, however, and it may be possible to use it to avoid harmful intersections. In fact, it is also not strictly necessary to use a \emph{single} surface: In \cite{Hirayama:1998ih} it is shown how to apply Stokes' Theorem to a collection of simply connected surfaces (some from a decomposition of a Seifert surface, some additional ones, it appears) to reconstruct the holonomy functional around a given knot.

\begin{figure}
\centerline{\epsfig{file=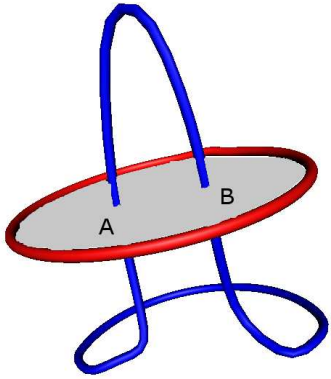, scale=0.27}}
\caption{\label{fi_two_circles2} The surface is intersected
twice by the link component. The intersection number is
still 0}
\end{figure}

\textbf{Consistency relations:} Due to the apparent freedom in the choice of the surfaces bounded by loops, there are possibly non-trivial consistency relations that one has to check. Consider for example the calculation of the expectation value of two unlinked unknotted loops. This can be done as described above, and leads to the right result. But one could also proceed differently and choose the surface $S$ bounded by the one loop such that it intersects the other. This situation is
sketched in figure \ref{fi_two_circles2}.
in that case one would have a non-trivial action of the surface-operator on the holonomy loop. One possibility is that the terms coming from the two intersections cancel each other in just the right way to yield the right expectation value. Another is that the loop ceases to be an eigenstate of the surface operator. In that case one would obtain relations between different expectation values. Many more situations exists in which the same expectation value can be obtained in different ways. These should be looked at.

Given the encouraging results of the present work, there are a number of questions that can be asked. For example, one should try to better understand 
what framing leads to the expectation values calculated here, since the trivial framing is not fully explaining the powers of $q$ that constitute the difference between our results and the results for standard framing. 

One should also ask why the Duflo isomorphism seems so important to get the CS expectation values right. One possible answer is, that since it is an isomorphism, the resulting ordering best respects the properties of the `classical' objects (the functional derivatives, or even the curvature) that they replace. But it could also be that the Duflo map is of more fundamental importance for CS theory. 

It would seem that the approach taken in this work exposes relations among the expectation values that may be not entirely obvious in other approaches. At least, it `explains' some such relations in terms of surfaces, intersections etc. Examples are the fact that all the calculated expectation values are proportional to the unknot, and the fact that expectation values of links consisting of unlinked components factorize. Thus one can ask if one can ultimately learn something new and non-trivial about CS theory in this way. 

Finally, given the fact that our method borrows some of the techniques of LQG, it can be asked if, vice versa, this method can be useful for LQG. Indeed, we have found that quantization using the Duflo map has very desirable properties. This may strengthen the case -- first laid out in \cite{Alekseev:2000hf} -- for applying it to the quantization of area in LQG. Moreover:
\begin{itemize}
	\item The surface operators are (path ordered) exponentials of  the surface operators from \cite{Thiemann:2000bv}. 
	\item The simplicity constraints in spinfoam formulations of gravity are imposed by turning variables into operators under the path integral in much the same way as was done here. Moreover, the operators are composed of invariant vector fields on Lie groups. Thus the Duflo map may be used to order them.
	\item The relation $F(A)\propto E$ shows up as the boundary condition on an isolated horizon \cite{Engle:2009vc}. This is yet another way to see why quantum CS theory is relevant for the theory of black hole horizons in loop quantum gravity. 
\end{itemize}
We hope to come back to these areas for potential applications of the techniques of this article in the future. 
\section*{Acknowledgments}
H.S.\ thanks L.\ Freidel for introducing him to the Duflo isomorphism. This research was partially supported by the Spanish MICINN project No.\ FIS2008-06078-C03-03.
\begin{appendix}
\section{Chern-Simons theory expectation values}
\label{ap_jones}
\begin{figure}
\centerline{\epsfig{file=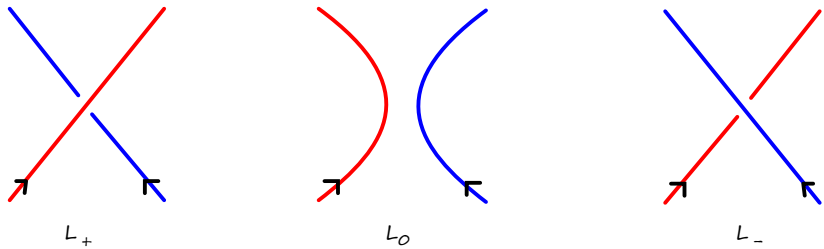, scale=0.45}}
\caption{\label{fi_skein} Neighborhood of a crossing with labels as used in the skein relation.}
\end{figure}

The SU(N) CS expectation values on $S^3$ and some more general manifolds have first been calculated by Witten in \cite{Witten:1988hf}, using tools from conformal quantum field theory. It is also in that reference that the connection between SU(2) CS theory and the  Jones polynomial is established. 
In our calculations, expectation values naturally come out in the normalization used in \cite{Witten:1988hf}, in which the the Jones polynomial
of unlinked knots factorizes. This normalization is characterized by the unknot expectation value 
\begin{equation}
\expec{\bigcirc}=\frac{q^{\frac{N}{2}}-q^{-\frac{N}{2}}}{q^{\frac{1}{2}}-q^{-\frac{1}{2}}}.
\end{equation}
Here, $q$ is a complex number of modulus one, that is related to the level $k$ of the CS theory by
\begin{equation}
\label{eq_qdef}
q=\exp\left(\frac{2\pi i}{N+k}\right).
\end{equation}
More generally, it was established in \cite{Witten:1988hf} that the expectation values for holonomy links in the defining representation are characterized by a skein relation, 
\begin{equation}
-q^{\frac{N}{2}}\expec{L_+} +(q^{\frac{1}{2}}-q^{-\frac{1}{2}})\expec{L_0}+q^{-\frac{N}{2}}\expec{L_-}=0
\end{equation}
where $L_+,L_0,L_-$ are holonomy links obtained from one another by changing the crossing structure at a crossing in a planar projection of the link (see figure \ref{fi_skein}). 
There are two subtleties in these results that we need to briefly discuss:
 
First off, the appearance of the shift by $+N$ in the denominator of the exponent in \eqref{eq_qdef} which was argued to be present in \cite{Witten:1988hf} is somewhat mysterious. It was discussed by various authors, and its presence is actually sometimes disputed (see for example \cite{Guadagnini:1994cc} for a discussion of this question).\footnote{In fact, it could be argued that the issue is somewhat moot, as the level $k$ appearing in the action would be the `bare' coupling constant, whereas whatever appears in expectation values is (related to) the `renormalized' coupling constant. The relation between the two would be dependent on the renormalization scheme and hence physically irrelevant. All that matters is the relationships between different expectation values.}  
In our approach we do not see this shift. Consequently, the $q$ appearing in the main text always denotes 
\begin{equation}
q=\exp\left(\frac{2\pi i}{k}\right).
\end{equation}
Secondly, it is well known that the CS expectation values actually depend on a choice of framing for the holonomy loops. This can be understood as the remnant of a covariant regularization procedure which renders the expectation values finite.   
If $L$ is a link with components $\{L_i\}$ and
\begin{equation}
\Delta=\sum_i \Delta\chi(L_i,L_i)
\end{equation}
the total change of self-linking, then the change of the expectation value is by a factor of
\begin{equation}
q^{\Delta h_R},
\end{equation}
where $h_R$ is a certain numerical constant depending on the structure group. 
For SU(N) it is given as 
\begin{equation}
h_R=\frac{N^2-1}{2N(N+k)}
\end{equation}
in \cite{Witten:1988hf}. We note again the shift by $N$. In our case, it is again natural to disregard this shift. Then we note 
\begin{equation}
h_R
=\begin{cases} 
\frac{3}{4} & \text{ for SU(2)}\\
\frac{4}{3} & \text{ for SU(3)} 
\end{cases}.
\end{equation}
In the present work, the expectation values for some links are recomputed for $G=$SU(2), SU(3). We finish this appendix by giving the expectation values that are also calculated in the main text according to \cite{Witten:1988hf}, in standard framing.

For multiple unlinked unknots: 
\begin{equation}
\expec{\underbrace{\bigcirc\ldots\bigcirc}_{n \text{ times}}}=\left(\frac{q^{\frac{N}{2}}+q^{-\frac{N}{2}}}{q^{\frac{1}{2}}-q^{-\frac{1}{2}}}\right)^n,
\end{equation}
For the two Hopf links (see figure \ref{fi_hopfpm}):
\begin{figure}
\centerline{\epsfig{file=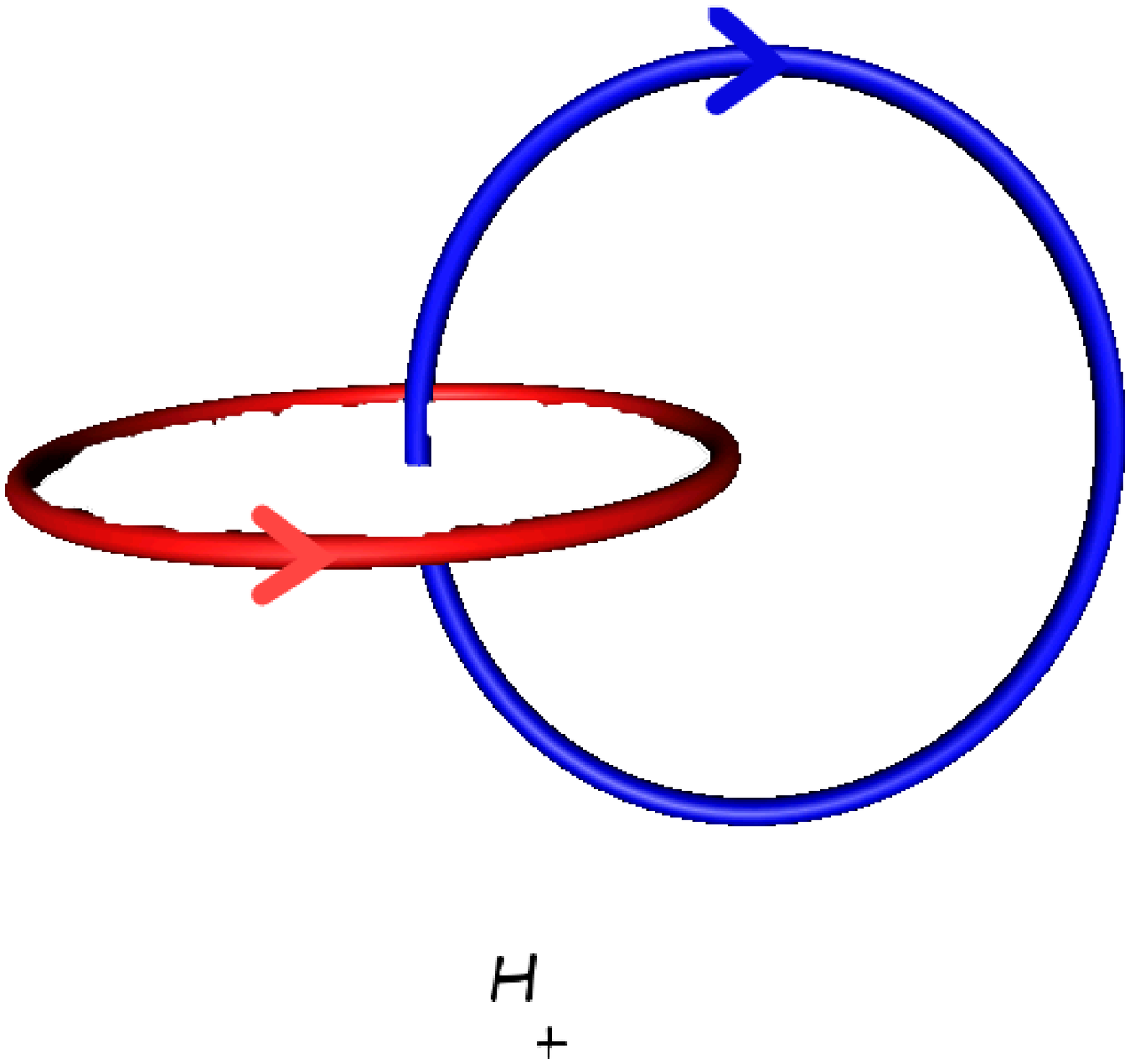, scale=0.25}$\,\,$\epsfig{file=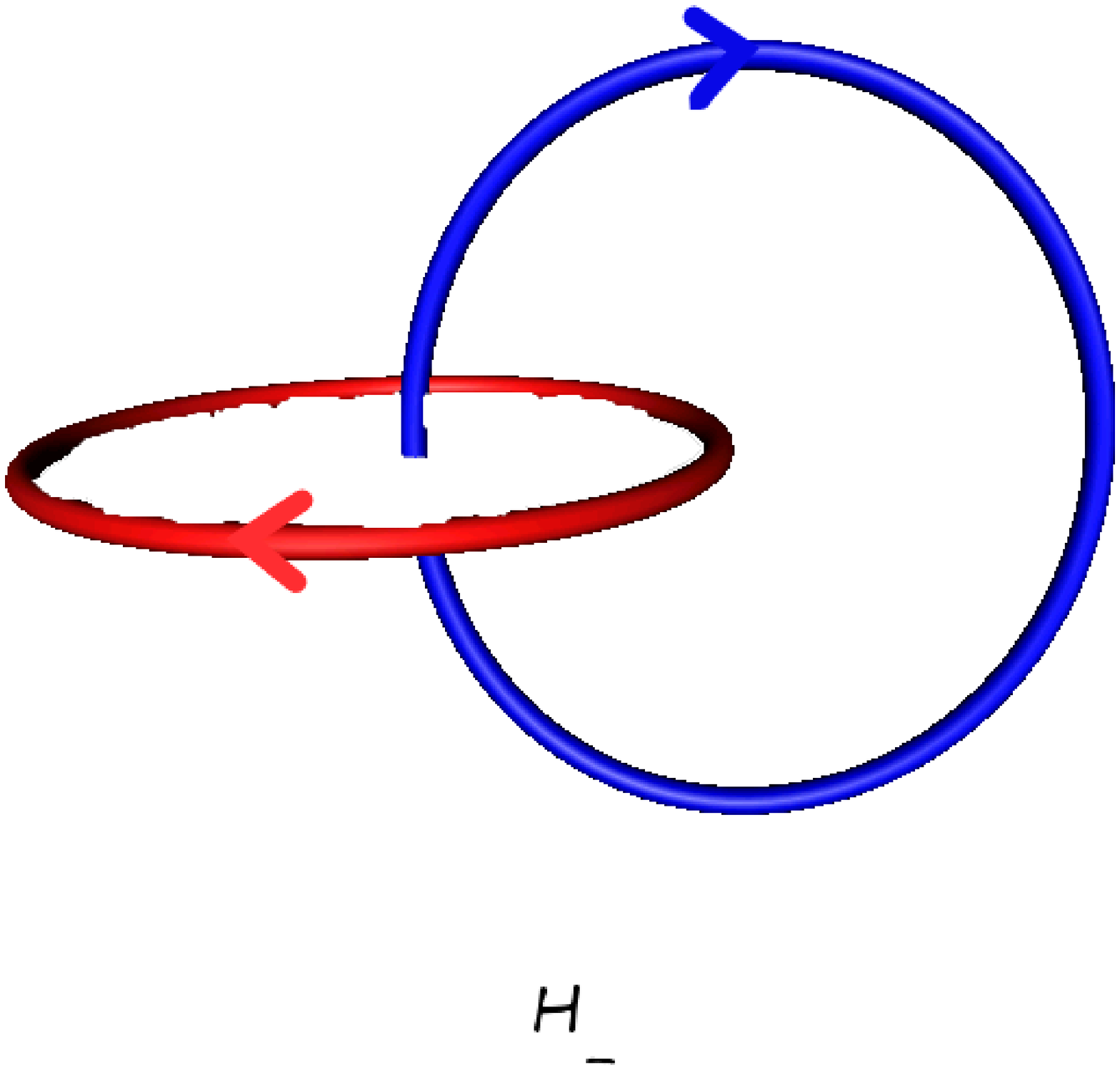, scale=0.25}}
\caption{\label{fi_hopfpm} The two Hopf links.}
\end{figure}
\begin{equation}\begin{split}
\label{eq_hopfright}
\expec{H_+}_{\SUTWO}
&=1+q^{-1}+q^{-2}+q^{-3}
=q^{-\frac{3}{2}}\expec{\bigcirc}_{\SUTWO}(q+q^{-1})\\
\expec{H_-}_{\SUTWO}
&=\overline{\expec{H_+}_{\SUTWO}}
=1+q+q^{2}+q^{3}=q^{\frac{3}{2}}\expec{\bigcirc}_{\SUTWO}(q+q^{-1})\\
\expec{H_+}_{\SUTHREE}
&=1+q^{-1}+2q^{-2}+2q^{-3}+2q^{-4}+q^{-5}
=\expec{\bigcirc}_{\SUTHREE}(q^{-1}+q^{-3}+q^{-4})\\
\expec{H_-}_{\SUTHREE}
&=\overline{\expec{H_+}_{\SUTHREE}}=1+q+2q^{2}+2q^{3}+2q^{4}+q^{5}
=\expec{\bigcirc}_{\SUTHREE}(q+q^{3}+q^{4})
\end{split}\end{equation}
For $n$ unknots that are linked to one unknot (see figure \ref{fi_keychain}): Let us denote that link by $L_n$. Then 
\begin{equation}
\expec{L_n}=q^{-\frac{N}{2}}(q^{\frac{1}{2}}-q^{-\frac{1}{2}})\expec{L_{n-1}}+q^{-n}\expec{\bigcirc}\expec{L_{n-1}}
\end{equation} 
For comparison with our results, it is also important to consider the effect of framing changes. The results above are obtained for standard framing, i.e., for the framing obtained from a Seifert surface. Inspecting the Seifert surface for the Hopf link from figure \ref{fi_seifert_example}, one concludes that both loops have self linking $+1$ in standard framing.  For the expectation values calculated in this article, it is very natural to assume that they will come out in a different framing, namely in the one given by discs bounded by the loops used in the calculations. In that framing, the self-linking is evidently 0. Let us therefore call this framing \emph{trivial}.
Thus we expect
\begin{equation}
\expec{H_+}=\expec{H_+}_{\text{trivial}}=q^{2h_R}\expec{H_+}_{\text{standard}} =\expec{H_+}_{\text{standard}}\cdot 
\begin{cases}
q^{\frac{3}{2}}& \text{ for SU(2)}\\
q^\frac{8}{3} & \text{ for SU(3)} 
\end{cases}
\end{equation}
where the expectation values on the very left are as calculated with the new method. Remarkably, this is exactly the case.  There seems to be a problem, however, with this interpretation in the case of the mirror Hopf-links. Since the self-linking should be independent of the orientation of the loop, one concludes that both loops have self linking $+1$ in standard framing also for $H_{-}$. Then 
\begin{equation}
\expec{H_-}_{\text{trivial}}=q^{2h_R}\expec{H_-}_{\text{standard}} =\expec{H_-}_{\text{standard}}\cdot 
\begin{cases}
q^{\frac{3}{2}}& \text{ for SU(2)}\\
q^\frac{8}{3} & \text{ for SU(3)} 
\end{cases}
\end{equation}
which is not what we find. So in this case we can interpret our result as being obtained in standard framing. 

For completeness, let us note that the connection between CS theory and the Jones polynomial of knot theory is as follows \cite{Witten:1988hf}:
The Jones polynomial is obtained by taking
CS expectation values of holonomy loops in the $j=1/2$ representation of SU(2), using standard framing. 

\section{Lie-algebra results}
\label{ap_lie}
For convenience and completeness, we list here some results regarding the defining and adjoint representation of SU(2) and SU(3) that are used in the main text. 

For an algebra element $x$ we have 
\begin{equation} 
|x|^2=c\norm{x}^2
 \text{ with } c
 =\begin{cases}
 \frac{1}{4} &\text{ for SU(2)}\\
 \frac{1}{6} &\text{ for SU(3)}
\end{cases}
\end{equation}
where $|x|^2=\tr(\pi(x)^2),\norm{x}^2=\tr(\ad(x)^2)$ are traces in the defining and the adjoint representation. The latter is the Cartan-Killing norm of $x$. 
We also find the following quadratic Casimir operator in the defining representation:
\begin{equation}
k^{IJ}T_IT_J
 =\begin{cases}
 \frac{3}{8}\um &\text{ for SU(2)}\\
 \frac{4}{9} \um &\text{ for SU(3)}
\end{cases}.
\end{equation}
For SU(3), the cubic Casimir is 
\begin{equation}
k^{II'}k^{JJ'}k^{KK'}\tr(T_IT_JT_K)T_{I'}T_{J'}T_{K'}=-\frac{2}{3^5}\um. 
\end{equation}
But the relevant quantity for the main text is actually the slightly different
\begin{equation}
k^{II'}k^{JJ'}k^{KK'}\tr(T_{(I}T_JT_{K)})T_{I'}T_{J'}T_{K'}=\frac{5}{2\cdot3^5}\um. 
\end{equation}
These results can certainly be obtained from the root and weight structure of the Lie algebras, but since we are not interested in generality, they are easiest to obtain by choosing a specific basis and calculating. For SU(3) one can choose for example the Gell Mann basis,
\begin{align*}
T_1&=\left(
\begin{array}{ccc}
 0 & \frac{1}{2} & 0 \\
 \frac{1}{2} & 0 & 0 \\
 0 & 0 & 0
\end{array}
\right),\qquad T_2&=\left(
\begin{array}{ccc}
 0 & -\frac{i}{2} & 0 \\
 \frac{i}{2} & 0 & 0 \\
 0 & 0 & 0
\end{array}
\right),\qquad T_3&=\left(
\begin{array}{ccc}
 \frac{1}{2} & 0 & 0 \\
 0 & -\frac{1}{2} & 0 \\
 0 & 0 & 0
\end{array}
\right),\\
\qquad T_4&=\left(
\begin{array}{ccc}
 0 & 0 & \frac{1}{2} \\
 0 & 0 & 0 \\
 \frac{1}{2} & 0 & 0
\end{array}
\right),\qquad T_5&=\left(
\begin{array}{ccc}
 0 & 0 & -\frac{i}{2} \\
 0 & 0 & 0 \\
 \frac{i}{2} & 0 & 0
\end{array}
\right),\qquad T_6&=\left(
\begin{array}{ccc}
 0 & 0 & 0 \\
 0 & 0 & \frac{1}{2} \\
 0 & \frac{1}{2} & 0
\end{array}
\right),\\
\qquad T_7&=\left(
\begin{array}{ccc}
 0 & 0 & 0 \\
 0 & 0 & -\frac{i}{2} \\
 0 & \frac{i}{2} & 0
\end{array}
\right),\qquad T_8&=\left(
\begin{array}{ccc}
 \frac{1}{2 \sqrt{3}} & 0 & 0 \\
 0 & \frac{1}{2 \sqrt{3}} & 0 \\
 0 & 0 & -\frac{1}{\sqrt{3}}
\end{array}
\right).
\end{align*}
In this basis, one finds 
\begin{equation}
\tr(T_I T_J)=\frac{1}{2} \delta_{IJ}, \qquad \tr(\ad(T_I) \ad(T_J))=3\delta_{IJ}
\end{equation}
and 
\begin{equation}
\sum_{IJK}\tr(T_IT_JT_K)T_{I}T_{J}T_{K}=-\frac{2}{9}\um, \qquad \sum_{IJK}\tr(T_{(I}T_JT_{K)})T_{I}T_{J}T_{K}=\frac{5}{18}.
\end{equation}

\end{appendix}

\end{document}